\pdfoutput=1
\documentclass[aps,pre,twocolumn,amsmath,amssymb,notitlepage,superscriptaddress,longbibliography,nofootinbib,10pt]{revtex4-1}
\usepackage[T1]{fontenc}
\usepackage[left]{lineno}
\usepackage{amsmath}
\usepackage{amssymb}
\usepackage{bm}
\usepackage{graphicx}
\usepackage{xcolor}
\definecolor{bleuf}{rgb}{0,0,1}
\definecolor{redf}{rgb}{0.61,0.24,0.38}
\usepackage[%
  colorlinks=true,
  urlcolor=redf,
  linkcolor=redf,
  citecolor=redf
]{hyperref}
\usepackage{url}
\usepackage{soul}
\usepackage{upgreek}
\usepackage[normalem]{ulem}
\usepackage{float}
\usepackage{appendix}

\begin{document}

\title{Phase-locking of hybrid oscillators}

\author{Joshua H. K. Saldi}
\affiliation{Huygens-Kamerlingh Onnes Laboratory, Universiteit Leiden,
PO Box 9504, 2300 RA Leiden, the Netherlands}
\author{Alexandre Morin}
\affiliation{Huygens-Kamerlingh Onnes Laboratory, Universiteit Leiden,
PO Box 9504, 2300 RA Leiden, the Netherlands}
\email{morin@physics.leidenuniv.nl}
\begin{abstract}
The synchronization and phase-locking behavior of oscillators with smooth dynamics is well captured by continuous phase-models {\em \`a la} Kuramoto.
However, these models do not apply seamlessly for hybrid oscillators, where discrete events occur along their otherwise smooth dynamics.
Consequently, the synchronization and phase-locking mechanisms of hybrid oscillators remain overlooked.
Here, we combine experiments, theory, and simulations, to investigate and rationalize the coherent motion of pairs of hybrid oscillator.
Using contact-charge electrophoretic (CCEP) oscillators as an experimental realization, we show that in-phase oscillations can occur, despite oscillators repelling each other.
We rationalize this behavior by introducing a discrete-time framework that explicitly accounts for discrete events and elucidates the origin of phase-locking.
Our model highlights the roles played by inertia and the decay of interaction strength along the oscillation cycle in promoting phase-locking of CCEP oscillators.
More generally, it provides a phase-locking criterion that holds for a broader class of coupled hybrid oscillators, and reveals that various mechanisms can lead to their coherence.
\end{abstract}

\maketitle

Rich dynamics emerge in active materials where oscillatory and spatial dynamic are coupled.
From the seminal swarmalators~\cite{o2017oscillators} and the pulsating active materials~\cite{zhang2023pulsating}, to living chiral crystals~\cite{chao2026selective}, such a bi-directional coupling not only underlies their collective behavior, but also enriches the range of collective states they can exhibit.
Indeed, in such systems, spatial organization can be controlled indirectly, by acting on the phase dynamics of the constituents~\cite{o2017oscillators,zhang2023pulsating,chao2026selective,le2025control}.
This novel possibility revives a classical challenge when dealing with populations of oscillators: understanding and modeling their synchronization and phase-locking behavior.

For a broad class of oscillatory systems, the solution is well-known.
In these systems, the dynamics can be expressed in terms of an internal phase-variable, or can be reduced to such a description~\cite{nakao2016phase}.
Within this framework, interactions between oscillators act by adjusting their relative phases.
The seminal Kuramoto model is an example of such a phase-reduced description~\cite{acebron2005kuramoto}.
It captures how weakly-coupled oscillators adjust their phases through simple interaction rules.
The model has been highly successful in describing synchronization phenomena in systems with smooth, continuous dynamics, such as arrays of mechanical oscillators, the coordinated beating of flagella, or the calling behavior of Japanese tree frogs~\cite{pantaleone2002synchronization,tuatulea2022elastohydrodynamic,brumley2014flagellar,aihara2009modeling}.

However, not all oscillators fit seamlessly into this framework.
In particular, oscillators could have their smooth dynamics disrupted, enabled, or even powered by discrete events.
An everyday example is a bipedal ramp-walker marching down an incline.
Other instances include the Mercury Beating Heart system, the integrate-and-fire neuronal models~\cite{lin1974mechanism,izhikevich2003simple}, and contact-charge electrophoretic (CCEP) oscillators~\cite{dou2018emergence,le2025control}.
By combining continuous motion with discontinuous events, the systems mentioned above constitute so-called \textit{hybrid oscillators}~\cite{lin2022hybrid}.
As their dynamics cannot be easily described with standard phase-reduction techniques, elucidating the emergence of synchronization in these systems remains a significant challenge~\cite{shirasaka2017phase}.

Here, we combine experiments, theory, and numerical simulations to investigate phase-locking of hybrid oscillators.
Our experimental setup is motivated by the collective phenomena reported in~\cite{le2025control} and inspired by the work of Eslami et al.~\cite{eslami2016modeling}.
It consists of CCEP oscillators - electrically-driven metallic sphere repeatedly colliding with one electrode.
Despite repelling each other electrostatically, we observed that these oscillators develop coherent, in-phase dynamics.
This is not expected from standard synchronization scenarios, where repulsive interactions typically favor anti-phase states~\cite{vathakkattil2020limits, strogatz2000kuramoto}.
To elucidate our experimental observations, we formulate a discrete-time model and derive a criterion for phase-locking.
The analysis reveals that inertia and decaying interactions promote the phase-locking observed experimentally.
Importantly, beyond these system-specific insights, our model applies to a broader class of coupled hybrid oscillators, including both underdamped and overdamped systems, with either repulsive or attractive couplings.

This Article is organized as follows.
In Section~\ref{Section2}, we investigate experimentally the dynamics of hybrid oscillators. 
We start by characterizing their single oscillator dynamics, and proceed with the two-body dynamics, which reveal phase-locking over a finite window of control parameters.
In Section~\ref{Section3}, we introduce a general discrete-time model and derive the criterion for phase-locking.
Unlike in our experiments, this model does not account for heterogeneities or noise.
In Section~\ref{Section3:Heterogeneities}, we demonstrate that detuned hybrid oscillators synchronize their periods.
We conclude by presenting a numerical phase diagram fully recapitulating our experimental observations.

\section{Experiments: Phase-locking dynamics of hybrid oscillators}\label{Section2}
\subsection{Single oscillator dynamics}
\begin{figure} [h]
\includegraphics[width=0.89\columnwidth]{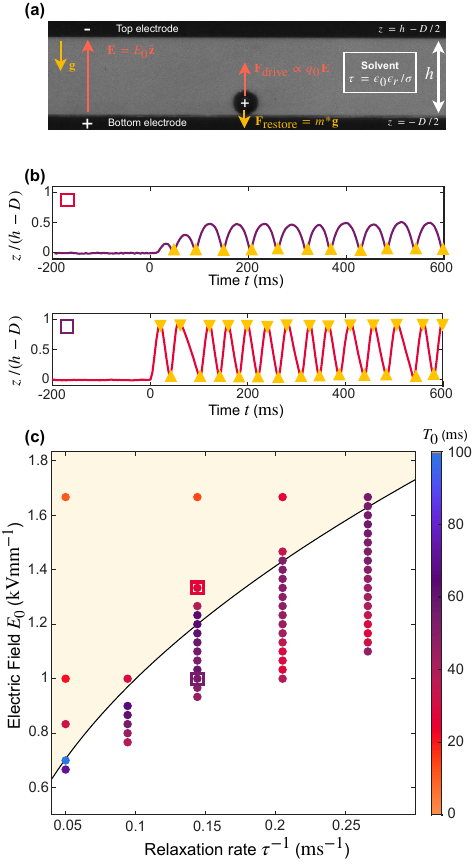}
\caption{
    {\bf Two regimes of spontaneous oscillations}
    {\bf (a).} Experimental snapshot showing a single contact-charge electrophoresis oscillator.
    A metallic bead between two electrodes is set into motion when the electric driving force $\mathbf{F}_{\rm drive}$ overcomes the restoring force of gravity $\mathbf{F}_{\rm restore}$.
    Two regimes of oscillations can be achieved by varying the amplitude of the electric field $E_0$ and the solvent charge-relaxation timescale $\tau$. $D=1\,\rm mm$ and $h=3\,\rm mm$.
    {\bf (b).} Experimental trajectories of the oscillator in the two regimes. Top: the particle reaches the top electrode and undergoes charge polarity switching. Bottom: charge relaxation limits the oscillation amplitude and the particle repeatedly collides with the bottom electrode only.
    Upward/downward triangles mark collisions with the bottom/top electrodes.
    {\bf (c).}
    Oscillation period $T_0$ of the oscillator for different values of $E_0$ and $\tau^{-1}$ (colored dots).
    The discontinuity of $T_0$ separates the two regimes of oscillation and coincides with the predicted scaling $E_0\propto\sqrt{h/\tau}$ with $\eta = 0.5$ (black line).
    }
    \label{Fig1}
\end{figure}

We first describe and characterize the single dynamics of the hybrid oscillators in our experiments.
Figure~\ref{Fig1}(a) shows a snapshot of the experimental setup: an aluminum sphere of mass $m=1.5\,\rm mg$ and diameter $D=1\,\rm mm$, confined between two electrodes spaced by a distance $h=3\,\rm mm$.
The sphere is turned into a CCEP oscillator moving between the electrodes by the application of an electric field~\cite{mersch2011antiphase, eslami2016modeling, drews2015contact}.
The oscillation mechanism is as follows:
When in contact with the electrode, the particle acquires a contact-charge $q_0$ proportional to the electric field $\mathbf{E}=E_0\hat{\mathbf{z}}$.
Therefore, it experiences an electric force $\mathbf{F}_{\rm drive}\propto q_0\mathbf{E}$ which attracts it towards the oppositely-charged electrode.
At high enough field amplitude $E_0$, the driving force overcomes gravity $\mathbf{F}_{\rm restore}=-mg\mathbf{\hat{z}}$ and the particle lifts off and sets into motion.

As shown in Fig.~\ref{Fig1}(b), the particle oscillates steadily after a short transient.
Importantly, under a Direct Current (DC) electric field, two qualitatively different regimes of spontaneous oscillations can be obtained:
Either the particle collides with both electrodes every cycle (top panel, see also Supplemental Movie 1), or it repeatedly collides with the bottom electrode only (bottom panel).
At the collisions with the electrodes, the particle is instantaneously recharged and undergoes an inelastic collision where its velocity $\dot{z}$ is updated through the rule $\dot{z}\rightarrow -\eta\dot{z}$.
Here, $\eta\in[0,1]$ is the restitution coefficient and characterizes the energy lost due to inelastic collision.
The combination of these discrete reset events and the continuous motion between them makes the CCEP oscillator a genuine hybrid dynamical system.

To achieve and select between the two different regimes of oscillation, we immersed the particle in a solvent and tuned its electrical conductivity $\sigma$ (see SM), as pioneered by Eslami et al.~\cite{eslami2016modeling}.
The electrical conductivity introduces a charge relaxation timescale $\tau = \epsilon_0\epsilon_r/\sigma$, where $\epsilon_0\epsilon_r$ is the permittivity of the medium.
When the particle departs from the electrode, its charge decays exponentially $q(t)=q_0e^{-t/\tau}$ due to Maxwell charge relaxation in the surrounding medium~\cite{Jackson:1998nia}.
As a result, $\mathbf{F}_{\rm drive}$ also decays.
If charge relaxation sufficiently suppresses $\mathbf{F}_{\rm drive}$ such that the maximum attained height $z_{\rm max}<h-D$, the particle does not reach the top electrode and it repeatedly collides with the bottom electrode.
On the other hand, if relaxation is slow such that $z_{\rm max}=h-D$, the particle collides with both electrodes during the cycle.

Importantly, the particle re-acquires its initial charge with opposite polarity $-q_0$ upon contact with the top electrode.
The electrical force, now pointing downwards, is restored for the second half of the oscillation cycle which reduces the oscillation period $T_0$.
As a consequence, the transition between the two regimes of oscillation is marked by a discontinuity in $T_0$.
This discontinuity is well-visible in Fig.~\ref{Fig1}(c), which shows the experimental phase-diagram established by systematically varying $E_0$ and $\tau^{-1}$ and measuring $T_0$.
Furthermore, the regime boundary follows the theoretical prediction $E_0\propto\sqrt{h/\tau}$, where the proportionality constant is determined by the restitution coefficient (see SM). 
A good agreement between theory and experiments is obtained with $\eta \simeq 0.5$.

With this thorough characterization of the single dynamics of CCEP oscillators, we now turn to investigating their binary behavior. 

\subsection{Two-body dynamics and phase-coherence}\label{Section4}
\begin{figure}[h]
\includegraphics[width=0.99\columnwidth]{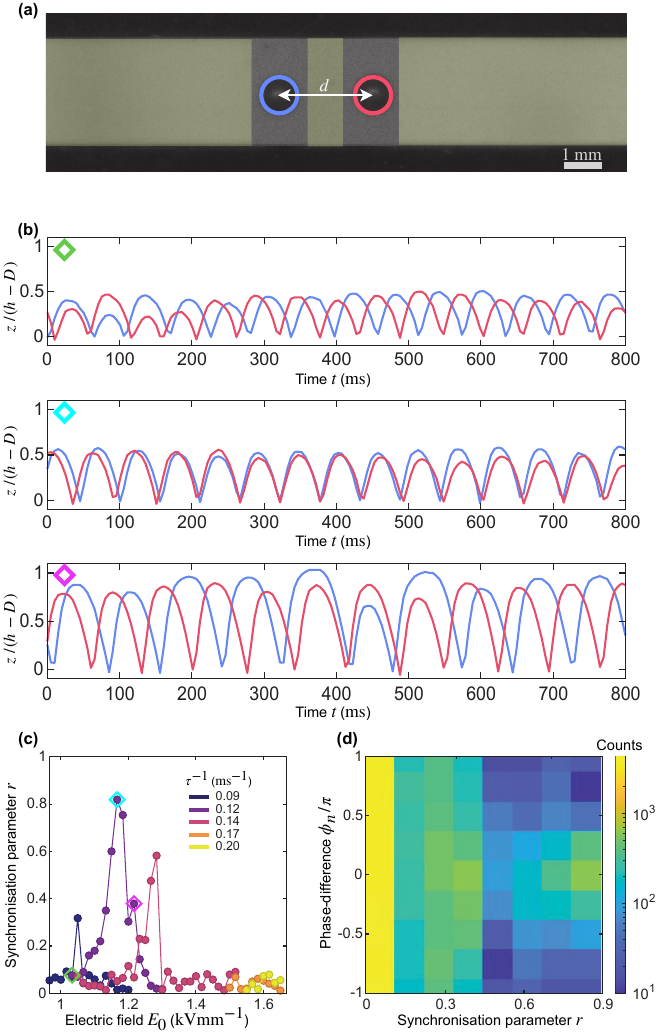}
\caption{
    {\bf Experimental evidence of synchronization dynamics}
    {\bf (a).} Experimental snapshot of two contact-charge electrophoretic oscillators. PMMA blocks maintaining a fixed distance $d=2.5\,\rm mm$ between the two oscillator are highlighted in yellow.
    {\bf (b).} Representative trajectories of two interacting oscillators for increasing values of the electric field $E_0$.
    In-phase motion occurs for intermediate values of $E_0$ only.
    {\bf (c).} Synchronization parameter $r$ as a function of $E_0$ for several values of $\tau^{-1}$.
    Synchronization occurs over a range of electric field which shifts to higher values as $\tau^{-1}$ increases.
    {\bf (d).} The bi-variate histogram of the synchronization parameter and phase-difference shows that synchrony corresponds to in-phase oscillations.
}
    \label{Fig5}
\end{figure}

We now build on this experimental realization of hybrid oscillators to investigate the emergence of synchronized motion between pairs.
In this Article, we confine ourselves to the regime where the CCEP oscillators only collide with the bottom electrode and do not switch charge polarity. 
We refer the reader interested in the coupled dynamics of CCEP oscillators with polarity switching to references~\cite{mersch2011antiphase, le2025control} where the limit $\tau\rightarrow \infty$ is investigated.

Figure~\ref{Fig5}(a) shows a snapshot of the experimental setup where two identical spheres are maintained at a fixed lateral distance $d$ by means of PMMA blocks.
Importantly, the PMMA wall does not merely set the distance between the two oscillators, it also removes hydrodynamic interactions between particles while preserving electrical interactions.

Typical trajectories for $\tau^{-1}=0.12\,\rm ms^{-1}$ are shown in Fig.~\ref{Fig5}(b) (see Supplemental Movies 2 and 3).
They reveal a qualitative and non-monotonous change in the two-body dynamics as the electric field amplitude increases.
At both low and high field, the dynamics is out-of-phase, while they are in-phase at intermediate field.

To quantify the synchronization behavior of these hybrid oscillators, it is advantageous to focus on their reset events rather than the continuous parts of their dynamics.
We therefore introduce the complex order-parameter:
\begin{equation} \label{Eq:OrderParameter}
    \boldsymbol{\Psi}_n=e^{2\pi i\Delta_n/T_0},
\end{equation}
where $\Delta_n$ is the event-time mismatch of the $n^{\rm th}$ oscillation cycle, that is, the duration between the collisions with the electrodes of the two particles.
The order parameter $\boldsymbol{\Psi}_n$ is a discrete variant of the one typically employed in the descriptions of continuous-phase oscillators, with the discrete phase-difference $\phi_n=\mathrm{arg}(\Psi_n)$.
The synchronization behavior is captured by the scalar synchronization parameter:
\begin{equation}
    r=\left|\langle\boldsymbol{\Psi}_n\rangle_n\right|,
\end{equation}
where the average is taken over all the reset events.
Perfectly synchronized dynamics corresponds to $r=1$, while $r=0$ reveals completely non coherent dynamics.

With these tools at our disposal, we now explore the two-body dynamics systematically by varying the electric field amplitude and the charge relaxation time in Fig.~\ref{Fig5}(c-d). 
Figure~\ref{Fig5}(c) confirms the qualitative behavior described in Fig.~\ref{Fig5}(b).
For a fixed value of $\tau^{-1}$, the synchronization parameter is a non-monotonous function of the electric field.
In addition, synchronization is never perfect, and occurs within a finite range of $E_0$ which shifts to higher values when the charge relaxation rate increases.
Figure~\ref{Fig5}(d) shows the bi-variate histogram of $r$ and $\phi_n$ where all the pairs of control parameters of Fig.~\ref{Fig5}(c) are included. 
For high value of $r$, the histogram is peaked at a phase-difference $\phi_n=0$; when synchronization occurs, the oscillators are always in-phase.

This last observation is somewhat non-intuitive.
Indeed, since the two oscillators have the same charge polarity, they repel electrostatically.
Their in-phase motion contrasts with Kuramoto models where repulsive couplings promote anti-phase synchronization.
In the next Section, we aim at resolving this apparent paradox and capturing theoretically our experimental observations. Specifically, we aim at elucidating: i) How does phase-locking emerge?  ii) Why is in-phase dynamics stable? iii) What are the necessary ingredients for phase-coherence? Is the inertial nature of the dynamics essential? What is the role played by charge decay? 

\section{Theory: discrete-time phase-locking}\label{Section3}

We now explain our experimental findings from Section~\ref{Section4} theoretically.
Our goal is not only to rationalize the phase-locking behavior of CCEP oscillators, but also to find out what the necessary ingredients for the phase-locking of generic hybrid oscillators are.
We therefore introduce a general formalism to describe coupled hybrid oscillators.
To do so, we introduce the following notations that connect the particular case of CCEP oscillators to our general model.
We let $\ddot{z}_{\rm drive}=\lvert\mathbf{F}_{\rm drive}(0)\rvert/m = q_0E_0/m$ denote the typical acceleration allowing the particle to leave its reference position $z=0$.
We denote by $\ddot{z}_{\rm restore}=m^\star\lvert g\rvert/m$ the restoring acceleration that brings the particle back to its reference position.

\subsection{Limit-cycle solution of a single oscillator}\label{Section3:Single}
We first consider a single generic hybrid oscillator driven by an exponentially relaxing force, whose equation of motion takes the form:
\begin{equation} \label{Eq:Motion}
    \ddot{z}(s) = \ddot{z}_{\rm drive}e^{-s/\tau}-\ddot{z}_{\rm restore},
\end{equation}
where $\tau$ is the relaxation time of the driving force, and $s$ is the time since the last reset event.
This continuous dynamics is complemented by events occurring when $z=0$, where the so-called {\em reset-map} is applied:
\begin{align}\label{Eq:ResetMap1}
    &s \rightarrow 0, \\ \label{Eq:ResetMap2}
    &\dot{z} \rightarrow -\eta\dot{z}.
\end{align}
Equation~\eqref{Eq:ResetMap1} resets the driving force to its maximal value (corresponding to the recharging $q\rightarrow q_0$ of CCEP oscillators).
In Eq.~\eqref{Eq:ResetMap2}, $\eta$ is the restitution coefficient which relates the pre-event and post-event velocities.
Inelastic collisions with $0 \leq \eta < 1$ are the only source of dissipation in the dynamics.

\begin{figure}[h]
 \includegraphics[width=\columnwidth]{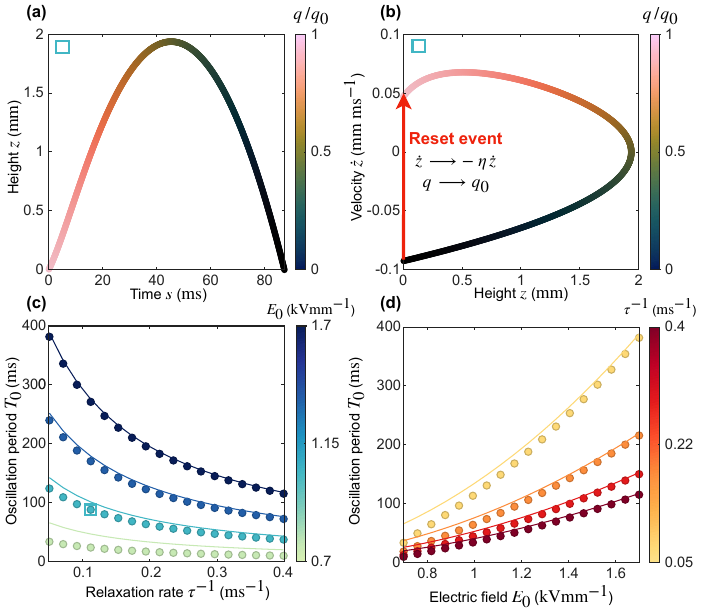}
\caption{
    {\bf Individual oscillator dynamics}
    {\bf (a).} Numerical trajectory of the periodic solution $z_0(t)$ to Eq.~\eqref{Eq:Motion} with discrete boundary conditions.
    We see that the solution is nearly parabolic, which resembles the experimental trajectories (see quenched oscillations in Fig.~\ref{Fig1}(b)).
    {\bf (b).} Phase-portrait of the hybrid limit-cycle.
    {\bf (c-d).} Numerical simulations of Eq.~\eqref{Eq:Motion} where we take $\ddot{z}_{\rm drive}\propto E_0^2$, inspired by our experimental setup from~\ref{Section2}.
    We extract the intrinsic oscillation period curves $T_0$ (coloured dots) for various values of $E_0$ and $\tau^{-1}$ and compare them against the analytical result Eq.~\eqref{Eq:Period} obtained from the periodic orbit of the oscillator dynamics (solid lines).
    For low values of $E_0$ and $\tau^{-1}$, there are some discrepancies between Eq.~\eqref{Eq:Period} and the simulations.
    However, for high values of $E_0$ and $\tau^{-1}$, the approximation $T_0\gg\tau$ becomes valid and the discrepancies disappear.
}
    \label{Fig2}
\end{figure}

Solving Eqs.~(\ref{Eq:Motion}-\ref{Eq:ResetMap2}) for a limit-cycle of period $T_0\gg\tau$ (see SM Section~{II.A}) yields:
\begin{equation} \label{Eq:Period}
T_0 = \frac{2}{1-\eta}\frac{\ddot{z}_{\rm drive}}{\ddot{z}_{\rm restore}}\tau.
\end{equation}
This shows that the relaxation time of the driving force sets the timescale of the hybrid oscillator.
Equation~\eqref{Eq:Period} is valid when the driving force overcomes the restoring force $\ddot{z}_{\rm drive} > \ddot{z}_{\rm restore}$ such that motion of the oscillator occurs in the first place.
Finally, if elastic collisions ($\eta=1$) occur during reset events, there is no dissipation along the dynamics and the period diverges.

We now confront these theoretical predictions to numerical solutions of Eqs.~(\ref{Eq:Motion}-\ref{Eq:ResetMap2}) in Fig.~\ref{Fig2} (see SM Section~{III} for numerical methods).
We show the limit-cycle trajectory and phase-portrait in Fig.~\ref{Fig2}(a-b) respectively for experimentally-relevant values of the control parameters.
The limit-cycle trajectory is almost parabolic, as expected when $T_0\gg\tau$.
More generally, Figs.~\ref{Fig2}(c-d) show that discrepancies in $T_0$ between theory and numerics remain limited to the lowest values of $E_0$, for which $\ddot{z}_{\rm drive} \simeq \ddot{z}_{\rm restore}$.
We recover the same trends as observed experimentally: the oscillation period decreases with the charge relaxation rate $\tau^{-1}$ and increases with $E_0$.

\subsection{Discrete-time phase-locking under weak coupling}\label{Section3:weakCoupling}
\begin{figure}[h]
 \includegraphics[width=\columnwidth]{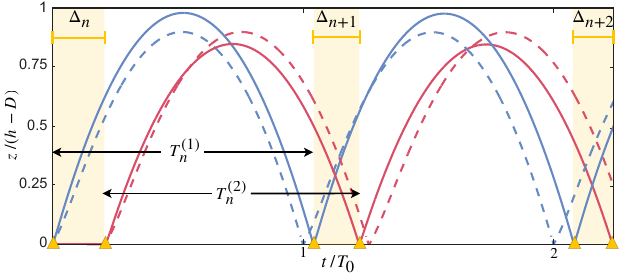}
\caption{
    {\bf Discrete-time description of weakly coupled oscillators}
    Numerical trajectories of two coupled hybrid oscillators (solid lines) and their intrinsic limit-cycle solution (dashed lines).
    The event-time mismatch $\Delta_n$ is calculated from the time-difference between the reset events of the oscillators (by the upward triangles).
    The coupling results in perturbed oscillation periods $T_n^{(1)}$ and $T_n^{(2)}$ of each of the oscillators.
    This leads to a decrease of $\Delta_n$ from cycle to cycle ($\Delta_{n}>\Delta_{n+1}>\Delta_{n+3})$, illustrating the contraction of event-time mismatches predicted by the recurrence relation Eq.~\eqref{Eq:ReturnMap}.
}
    \label{Fig:BinaryTrajectory}
\end{figure}

We now aim at elucidating the conditions enabling phase-locking between two oscillators.
We consider a pair of identical hybrid oscillators whose equation of motion reads:
\begin{equation}\label{Eq:MotionInteraction}
    \ddot{z}(s)=\ddot{z}_{\rm drive}e^{-s/\tau}-\ddot{z}_{\rm restore}+\varepsilon g_{\rm int}(z,s,\Delta_n),
\end{equation}
where we decomposed the interaction force into a dimensionless prefactor $\varepsilon$ and a term $g_{\rm int}$ of order $\ddot{z}_{\rm drive}$.
Here, $g_{\rm int}$ is a function of the height of the oscillator $z$, of the cycle-time $s$, and of the height of the other oscillator implicitly given by $s$ and the event-time mismatch $\Delta_n$.
As illustrated in Fig.~\ref{Fig:BinaryTrajectory}, $\Delta_n$ is the duration between the $n^{\rm th}$ reset events of the two particles.

As in Section~\ref{Section2}, we focus on these event-time mismatch to define phase-locking between the two oscillators, through the order parameter \eqref{Eq:OrderParameter}.
Due to the coupling, the oscillation period need not be constant cycle-to-cycle, and the evolution of $\Delta_n$ is given by
\begin{equation}\label{Eq:PhaseDifference}
    \Delta_{n+1}=\Delta_n+T_n^{(2)}-T_n^{(1)},
\end{equation}
where $T_n^{(1)}$ and $T_n^{(2)}$ are the perturbed oscillation periods due to the coupling between the oscillators, see Fig.~\ref{Fig:BinaryTrajectory}.

We determine $T_n^{(1)}$ and $T_n^{(2)}$ in the Supplemental Material. 
The derivation relies on an expansion around the limit-cycle solution of the uncoupled oscillator to first-order in $\varepsilon$, and the adjoint function to obtain closed-form expressions (see SM Section~{II.B} and refs.~\cite{shirasaka2017phase, luchini2024introduction}). 
The evolution of $\Delta_n$ takes the form of the recurrence relation
\begin{equation} \label{Eq:ReturnMap}
    \Delta_{n+1} = \Lambda\Delta_n.
\end{equation}
Before discussing the expression of $\Lambda$, a comment is in order.
Equation~\eqref{Eq:ReturnMap} shows that the synchronization dynamics reduce to a linear recurrence for the event-time mismatch.
It is the discrete-time analogue of the phase-difference equation in Kuramoto-type theories.
Rather than continuously evolving phases, it governs the cycle-to-cycle evolution of event-time mismatches.

The recurrence relation~\eqref{Eq:ReturnMap} shows that the only possible coherent dynamics is when the oscillators move in-phase, admitting a single fixed-point $\Delta^*=0$.
But phase-locking effectively takes place if and only if this fixed-point is stable:
\begin{equation}\label{Eq:StabilityCriterion}
    \lvert\Lambda\rvert<1.
\end{equation}
When this phase-locking criterion is satisfied, event-time mismatches contract from one cycle to the next, leading to phase-locking.
In contrast, when $|\Lambda|>1$, the mismatches expand and phase-locking is not possible.

The contraction factor $\Lambda$ is given by
\begin{equation} \label{Eq:ContractionFactor}
    \Lambda=1-2\varepsilon K_0,
\end{equation}
where $K_0$ is the linear-response integral which measures the event-time mismatch produced by coupling during the elapsed cycle.
The full expression for the linear-response integral is given in Appendix~\ref{App:A} (see also SM Section~{II.B}).
Here, we focus on its simplified expression, corresponding to $\eta=0$, which captures the main physics behind phase-locking in a more illustrative way:
\begin{equation}\label{Eq:LinearResponseIntegral_noMem}
    K_0=\int_0^{T_0}ds\,\,\chi(s)\Gamma(z,s),
\end{equation}
where $\chi$ is a response function that only depends on the single oscillator dynamics and $\Gamma(z,s)$ is the interaction strength linearized around phase-locking:
\begin{equation}\label{Eq:Gamma}
    \Gamma(z,s)=\partial_{\Delta}g_{\rm int}(z,s,\Delta)\big\rvert_{\Delta=0}+\mathcal{O}(\Delta^2).
\end{equation}
Equation~\eqref{Eq:LinearResponseIntegral_noMem} has a simple interpretation: At any point $(z,s)$ during its cycle, the oscillator feels the perturbation $\Gamma$ caused by interactions. 
These perturbations cause a shift in the subsequent event-time mismatch.
How exactly the mismatch is shifted, does not only depend on the how large this perturbation is, but also on when during the cycle it acted.
The role of the response function $\chi(s)$ is to determine how much such an instantaneous perturbation acting at time $s$ affects the next event-time mismatch.
The total mismatch results from the accumulation of these instantaneous contributions along the cycle.

We observed in the experiments of Section~\ref{Section4} that the coherent dynamics of CCEP oscillators takes the form of in-phase motion.
Our analysis above shows that the phase-locking behavior of weakly-coupled hybrid oscillators is determined by the response integral~\eqref{Eq:LinearResponseIntegral_noMem}, which can be evaluated from the equation of motion.
In the next Section, we apply this formalism to CCEP oscillators and identify the necessary ingredients for in-phase motion observed in our experiments.

\subsection{Phase-locking of repulsive hybrid oscillators}\label{Section3:Coulomb}

We start with the expression for the response function $\chi(s)$, which is entirely determined by the adjoint formulation of the uncoupled oscillator dynamics from Section~\ref{Section3:Single} (see also SM Section~{II~B}).
It reads
\begin{equation}\label{Eq:ResponseFunction}
    \chi(s)=\frac{1-\eta}{\ddot{z}_{\rm drive}}\frac{T_0-s}{\tau},
\end{equation}
where $\ddot{z}_{\rm drive}=q_0E_0/m$ is the driving acceleration at the beginning of the cycle.
Importantly, $\chi$ is always positive and decreases along the cycle.
It shows that perturbations have a bigger effect on the contraction factor the earlier they act along the cycle.
This bias towards early-cycle perturbations is a consequence of the inertial nature of the equation of motion Eq.~\eqref{Eq:Motion} (see also SM Section~{II~C} and Discussion).

We continue with the expression for the interaction strength resulting from Coulomb repulsion between two oscillators:
\begin{align}
    & \varepsilon= \frac{1}{4\pi\epsilon_0\epsilon_r}\frac{q_0^2}{md^2\ddot{z}_{\rm drive}},\label{Eq:eps}\\
    & \Gamma(s)=\frac{\ddot{z}_{\rm drive}}{d}\dot{z}(s)e^{-2s/\tau} \label{Eq:CoulombInteraction}.
\end{align}
Here, $\dot{z}(s)$ is the limit-cycle velocity of the uncoupled oscillator from Section~\ref{Section3:Single}, $\epsilon_0$ is the vacuum permittivity, and $\epsilon_r$ is the relative permittivity of the surrounding medium.
The contact charge of CCEP oscillators is proportional to the electric field $\ddot{z}_{\rm drive}\propto E_0^2$, therefore $\varepsilon$ is independent of the experimental control parameters.
Since $\varepsilon\sim10^{-5}$, the weak coupling assumption is verified.
The inspection of $\Gamma(s)$ is enlightening when it comes to the origin of phase-locking.
Its amplitude decays exponentially due to charge relaxation.
This means that interactions are strongly localized near the beginning of the cycle, where $\dot{z}(s)$ is positive.

Taken together, the early-cycle biases of both $\chi$ and $\Gamma$ merge in the response integral $K_0$.
This is confirmed in the limit $T_0\gg\tau$ where $K_0$ scales as:
\begin{equation}
    \label{Eq:K_Coulomb_Scaling}
    K_0\propto \frac{1}{d}E_0^4\tau^2.
\end{equation}
The temporal weighting of perturbations discussed above is reflected in the sign of $K_0$, which is positive for all values of $E_0$ and $\tau^{-1}$.
Writing the phase-locking criterion Eq.~\eqref{Eq:StabilityCriterion} in terms of $K_0$ gives
\begin{equation}\label{Eq:KCriterion}
    0<K_0<\frac{1}{\varepsilon}.
\end{equation}
This shows that the hybrid oscillators can be brought into a phase-locked state by finely controlling the experimental control parameters.
The upper-bound of the condition Eq.~\eqref{Eq:KCriterion} determines when coupling starts to expand event-time mismatches, leading to the loss of phase-locking.
The shift of the boundary location towards larger values of $E_0$ for increasing values of $\tau^{-1}$ as predicted by Eq.~\eqref{Eq:K_Coulomb_Scaling} is consistent with the experimental observations of Fig.~\ref{Fig5}(c).

To sum up, the phase-locking behavior of the CCEP oscillators is made possible by two ingredients: i) the bias in sensitivity to early-cycle perturbations due to inertia, and ii) the charge relaxation which makes early-cycle interactions dominant.
The combination of these two effects leads to an effective restoring force on the event-time mismatches, even though the coupling remains purely repulsive in space.

\section{Synchrony and phase behavior}\label{Section3:Heterogeneities}
While the theory captures the loss of phase-locking at high electric field $E_0$ and low charge relaxation rate $\tau^{-1}$ (see Eqs.~\eqref{Eq:K_Coulomb_Scaling}-\eqref{Eq:KCriterion}), it does not account for the loss of coherence observed experimentally at low $E_0$ and high $\tau^{-1}$ (see Fig.~\ref{Fig5}(c)).
In experiments, imperfections and noise are unavoidable. 
In particular, oscillators can have slightly different intrinsic oscillation periods.
Our current theory, however, assumes identical oscillators and cannot account for differences in oscillation period.
This raises the question of the synchronization of detuned oscillators, which is a prerequisite for phase-locking.
We show in the Supplemental Material (Section~{II~D}) that in the linear regime, such two oscillators synchronize their oscillation periods, and that the phase-locking criterion Eq.~\ref{Eq:KCriterion} is unchanged.
However, the fixed-point $\Delta^\star$ is changed and increases with the difference in oscillation periods $\Delta T$ as
\begin{equation}\label{Eq:shiftedFixedPoint}
    \Delta^*=\frac{\Delta T}{1-\Lambda},
\end{equation}
where $\Lambda$ is the contraction factor (see Eq.~\eqref{Eq:ContractionFactor}).
While the linear theory always predicts synchrony, we note that since $\Lambda \rightarrow 1$ at low $E_0$ and high $\tau^{-1}$. 
Therefore, $\Delta^\star \rightarrow \infty$ in these limits, which breaks the small-$\Delta$ assumption Eq.~\eqref{Eq:Gamma}.

\begin{figure}[h]
\includegraphics[width=0.8\columnwidth]{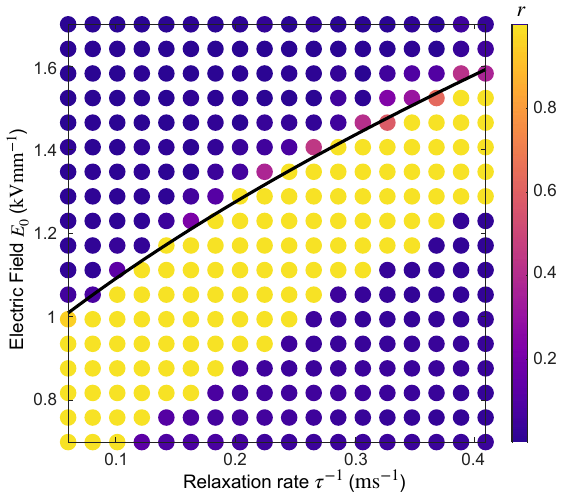}
\caption{
    {\bf Phase-diagram of synchronization} 
    Numerical simulations taken of the synchronization parameter $r$ for oscillators with a difference in oscillation period $\Delta T/T_0=10\,\rm \%$ interacting through Coulomb repulsion. The solid black line represents the upper bound in the phase-locking criterion Eq.~\eqref{Eq:KCriterion}.
}
    \label{Fig4}
\end{figure}

To conclude, we explore beyond the linear regime by solving the coupled dynamics of detuned oscillators numerically.
We tweak the masses of the two oscillators to implement a period difference $\Delta T/T_0 = 10\,\rm \%$, which are the typical fluctuations in the intrinsic oscillation periods measured experimentally (see SM Section~{I~C}).
The resulting phase-diagram is shown in Fig.~\ref{Fig4}.
In addition to the transition line corresponding to the upper-bound of the phase-locking criterion, the phase-diagram features another transition line at low $E_0$.
In this region, the coupling strength is too low to compensate for the intrinsic heterogeneities, and synchronization, hence phase-locking, fails.
Overall, our theoretical and numerical findings capture the experimental observations of phase-locking over a finite-range of control parameters.

\section*{Discussion}
By combining experiments, theory, and numerical simulations, we investigated the phase-locking behavior of hybrid oscillators.
Our analysis uncovers the origin of phase-locking observed in our experimental realization: electrically-coupled contact-charge electrophoretic oscillators undergoing charge relaxation.
The inertial dynamics and the decrease of interaction strength along the cycle caused by charge decay conspire to reduce the timing-mismatch of their reset events.
Together these two ingredients lead to in-phase motion.

Crucially, these ingredients follow from a generic rule predicting the phase-locking dynamics of any pair of coupled hybrid oscillators obeying Eq.~\eqref{Eq:MotionInteraction}.
This rule is expressed in terms of a response integral $K_0$, and phase-locking is predicted within a certain range of $K_0$.
This response integral combines how a single oscillator responds to perturbation with the interaction strength experienced along the cycle.
Therefore, our analysis suggests the existence of many pathways to phase-locking for hybrid oscillators.
We delve into such pathways in the SM Section~{II~C} and show that phase-locking is also possible without inertia, or without decaying interactions.
Finally, we rationalize the counter-intuitive observation that repulsive interactions between CCEP oscillators promote their in-phase motion.
Indeed, our analysis show that repulsive interactions are beneficial to phase-locking when they act in the ascending first half of the cycle, and are detrimental in the descending second half.
Attractive interactions have the complementary effect.

Our findings open several directions for future work.
First, the introduced framework provides a natural starting point to study synchronization in larger assemblies of hybrid oscillators.
Here, a discrete-time approach would advantageously capture phase-wave propagation resulting from the interplay between pairwise synchronization and spatial organization.
Second, the experimental realization could be extended to test the influence of the nature of the coupling predicted in theory, {e.g.} by using magnetic particles.
This would allow the coupling strength to be tuned independently from the active strength, and may provide additional strategies to control synchronization in active oscillatory matter.
Finally, our work shows that coherent oscillator dynamics can emerge without imposing a global phase-reset~\cite{le2025control} or constraining the system topology~\cite{baconnier2022selective}.
With synchronization arising directly from the internal dynamics, the system remains fully autonomous while maintaining stable internal coherence.
The experimental system therefore provides a promising platform to study collective excitations in oscillatory active systems.

\section*{Acknowledgments}
We thank Raphaël Zwier and Robin Schrama for help with the experimental setup. We thank Emanuel F. Teixeira for insightful discussions.

\section*{Author Contributions}
A.M. designed the project. J.H.K.S. performed the experiments, simulations, and theoretical analysis.
All authors discussed the results and wrote the manuscript. 
Correspondence to \href{mailto:morin@physics.leidenuniv.nl}{Alexandre Morin}.

\section*{Competing Interest}
The Authors declare no competing interests.
\section*{Data Availability}
All data supporting this study are available from the authors upon reasonable request.

\onecolumngrid
\appendix
% \appendixpage
\section{Analytical expressions of the linear-response integral}\label{App:A}
The full linear-response integral for an arbitrary finite $\eta<1$ appearing in the contraction factor Eq.~\eqref{Eq:ContractionFactor} reads:
\begin{align}
    &K = \mathcal{N}(\Pi,\eta)\left( K_0-\frac{\eta}{\eta+1}K_{\rm reset} \right), \label{EqS:I_mem}\\
    &K_{\rm reset}=T_0\chi(T_0)\big\langle\Gamma(s)\big\rangle_{T_0}, \label{EqS:K_reset}\\
    &\mathcal{N}(\Pi,\eta)=\frac{1+\eta}{1-\eta}\frac{1}{1+\eta\Pi e^{-\Pi}}.
\end{align}
where $K_0$ is the same expression as in Eq.~\eqref{Eq:LinearResponseIntegral_noMem} and $\Pi=T_0/\tau$.
For Coulomb coupling, the final expressions read
\begin{align}
    &K_0=\frac{\ell_{\rm drive}}{d}\mathcal{K}_0(\xi, \Pi),\\
    &K_{\rm reset}=\frac{\ell_{\rm drive}}{d}\mathcal{K}_{\rm reset}(\xi, \Pi),
\end{align}
where $\ell_{\rm drive} = \ddot{z}_{\rm drive}\tau^2$.
Finally, $\mathcal{K}_0$ and $\mathcal{K}_{\rm reset}$ are functions of the dimensionless parameters $\Pi$ and $\xi=\ddot{z}_{\rm drive}/\ddot{z}_{\rm restore}$:
\begin{align}
    &\mathcal{K}_0(\xi,\Pi)=+\frac{1}{4}(2\Pi-1+e^{-2\Pi})-\frac{2\xi}{9\Pi}(3\Pi-1+e^{-3\Pi})-\frac{1}{2\Pi}\left(\Pi-1+(\Pi+1)e^{-2\Pi}\right),\\
    &\mathcal{K}_{\rm reset}(\xi,\Pi)= +\frac{\Pi}{2}(1-e^{-2\Pi})-\frac{2\xi}{3}(1-e^{-3\Pi})-\frac{1}{2\xi}(1-e^{-2\Pi}(1-e^{-2\Pi})).
\end{align}

\twocolumngrid

\bibliography{biblio}
\end{document}